\documentclass{aa}
\usepackage{psfig}
\begin{document}

   \title{Two new bright Ae stars
   \thanks{Based on observations collected at the 1\,m telescope of
   the Special Astrophysical Observatory (Nizhnij Arkhyz, Russia)}}

   \author{D.N.Monin \inst{1,2}, G.A.Wade \inst{3}, S.N.Fabrika \inst{2}}

   \offprints{D.N.Monin, \email{dmonin@phobos.astro.uwo.ca}}

   \institute {Physics \& Astronomy Department,
   The University of Western Ontario, London, Ontario, N6A 3K7 Canada
   \and       Special Astrophysical Observatory of Russian AS,
   Nizhnij Arkhyz 369167, Russia
   \and       Department of Physics,
   Royal Military College of Canada, Kingston, Ontario, K7K 7B4 Canada}

   \date{Received / Accepted}

\abstract{
Two newly identified Ae stars, $\nu$\,Cyg and $\kappa$\,UMa, were discovered in the course of the Magnetic Survey of Bright MS stars (Monin et al.~2002). We present their H$\alpha$ profiles along with measurements of their equivalent width and parameters of emission features. Emission in the H$\alpha$ line of $\nu$\,Cyg is variable on a time scale of 3 years. $\kappa$\,UMa exhibits weak emission which is rather stable. The emission is thought to arise from a circumstellar disk, and we have estimated the size of that disk.Both new emission stars are IRAS sources. Their IR color excesses are consistent with those of classical Ae stars. Thus, $\nu$\,Cyg and $\kappa$\,UMa appear not to belong to the class of Herbig Ae/Be stars. We argue that the frequency of Ae stars may be underestimated due to the difficulty of detection of weak emission in some A stars.
      \keywords{stars: emission line, Ae -- stars: binary -- stars:
      Magnetic Survey:
      individual: $\nu$\,Cyg -- HD199629 -- HR8028:
      individual: $\kappa$\,UMa -- HD77327 -- HR3594}
}

   \maketitle
%
\section{Introduction}

Ae (A--emission) stars are believed to be cooler 
counterparts of classical Be stars (Jaschek and Andrillat~1998).
They have similar observational properties: emission in hydrogen lines,
IR excess compared to the IR brightness expected from the stellar photosphere,
photometric and spectral variability on timescales from days to years, and
linear polarization. In general, the phenomena that are observed in Ae 
stars are less spectacular then, but similar to, Be phenomena. One example
is the behavior of emission features. Be stars exhibit emission in different
spectral lines. The number of lines in emission 
and its strength decreases when one goes from late Be to early Ae stars
(Jaschek, Andrillat and Jaschek~1991). In stars later then A0, only 
H$\alpha$ and sometimes $H\beta$ are seen in emission (Jaschek et al.~1980).

The group of Ae stars is almost 10 times less numerous than the group of 
the hotter Be counterparts (Zorec and Briot~1997).
To enhance the probability of finding such stars, different 
selection criteria are used, for example, anomalous IR emission 
(Jaschek et al.~1991), variability (Irvine~1979), or fast rotation 
(Ghosh, Apparao and Pukalenthi~1999).

In the course of the Magnetic Survey of Bright Main Sequence stars 
(Monin et al.~2002) we obtained  high
resolution spectropolarimetry of 21 A type stars with visual magnitude $m_V<4$. 
There was only one known Ae star in our sample, $\gamma$\,UMa, 
indicated as an Ae star in 
the Bright Star Catalogue (BSC; Hoffleit and Jaschek~1982);
it did not show detectable emission in our observations.
However, inspection of Stokes I profiles of 21 A type stars
resulted in the discovery two new Ae stars, $\nu$\,Cyg and $\kappa$\,UMa.
In this paper we discuss these stars.

\section{The data used}

The Magnetic Survey of Bright Main Sequence Stars is being carried out using
the coud\'e \'echelle spectrograph CEGS of the 1m telescope of the Special Astrophysical Observatory of the Russian Academy of Sciences (SAO RAS).
The spectra cover the range from 4000 to 9000\,\AA, with a resolving power
of $R\simeq 40000$. The main goal of this project is to obtain high quality polarimetric spectra in order to 
detect new weakly magnetic stars among both rapidly and slowly rotating
main sequence stars. As a byproduct of these observations, we obtain very high signal-to-noise Stokes $I$ profiles which,
because of their very precise residual intensities,
are suitable for other purposes. During reduction, we pay special attention to the
spectral normalization especially in orders containing broad hydrogen lines 
(see description of the data reduction procedure by Monin et al.~2002).
Thus the Stokes $I$ spectrum is suitable, for example, for comparison
with synthetic spectra, and, because of its high spectral resolution, 
for a searches for and detailed study of weak emission features (which are generally undetectable at low resolution).

Two new Ae stars, $\nu$\,Cyg and $\kappa$\,UMa, were found among the targets
of the Magnetic Survey. The basic observational properties of these stars are listed in Table\ref{stars}.
\begin{table}
  \caption[]{Basic observational properties (according to the BSC) of the two bright Ae stars discovered during the Magnetic Survey.}
  \label{stars}
  \begin{tabular}{rrrccrr}
  \hline
           Name &   HR &     HD &   V  &  B-V &   Sp     & $v\sin i$, \\
                &      &        &      &      &          &    km/s    \\
            \hline                      
  $\kappa$\,UMa & 3594 &  77327 & 3.60 & 0.00 & A1Vn     & 219        \\
     $\nu$\,Cyg & 8028 & 199629 & 3.94 & 0.02 & A1Vn     & 241        \\
  \hline
  \end{tabular}
  \end{table}
Below we discuss them in detail.

\section{New Ae stars}\label{Ae}

\begin{figure}
  \hbox{
  \psfig{figure=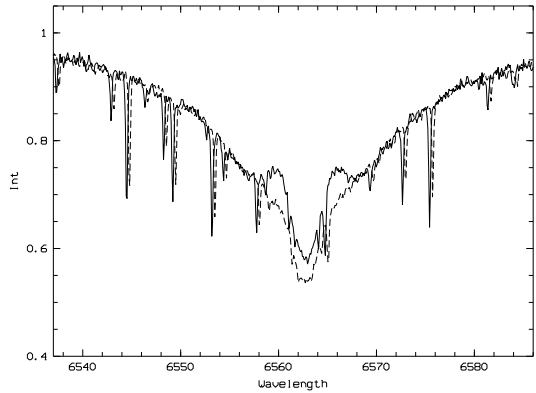,width=8cm,angle=270}}
  \parbox[l]{8cm}{ 
  \caption{H$\alpha$ profiles of $\nu$\,Cyg obtained in 1996 
  (solid line) and in 1999 (dashed line).
  The spectra have corrected to a heliocentric reference frame. 
  The sharp lines are telluric absorption lines.}
  \label{nuCyg_Halp}}
\end{figure}

Fig.~\ref{nuCyg_Halp} shows H$\alpha$ profiles in spectra of $\nu$\,Cyg obtained during 1996 and 1999.
The spectra have been shifted to the heliocentric reference frame, 
but have not been corrected for the radial velocity of the star.
A double-peaked emission inside the absorption H$\alpha$ core
is clearly seen in the 1996 data.
In 1999 H$\alpha$ had an absorption profile, but the broadening
does not seem consistent with the high rotational velocity of 240~km/s listed
in the BSC. This apparently low $v\sin i$ may well be a consequence of infilling of the line core by undetected emission.

\begin{figure}
  \vbox{
  \psfig{figure=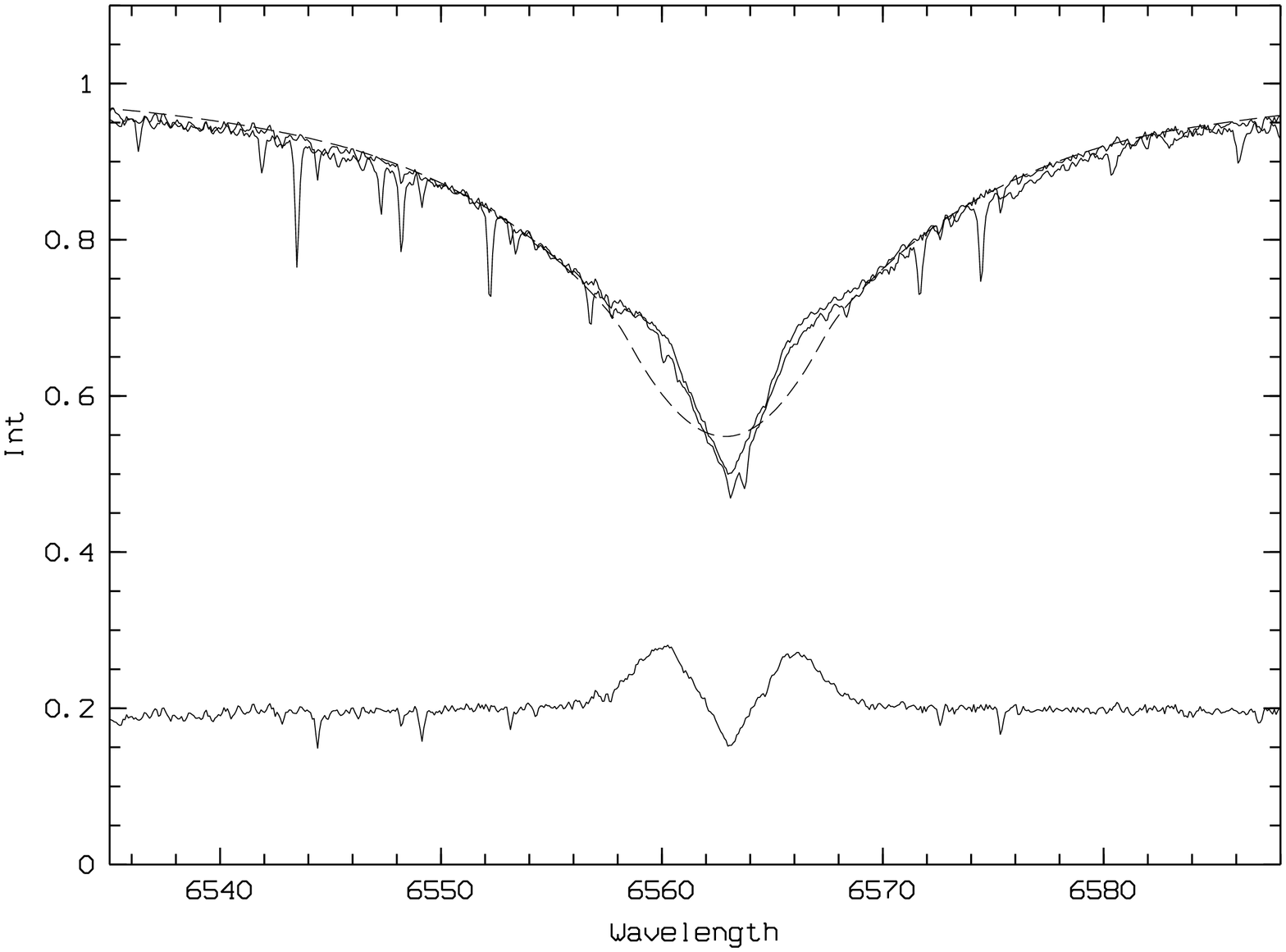,width=8cm,angle=270}
  \psfig{figure=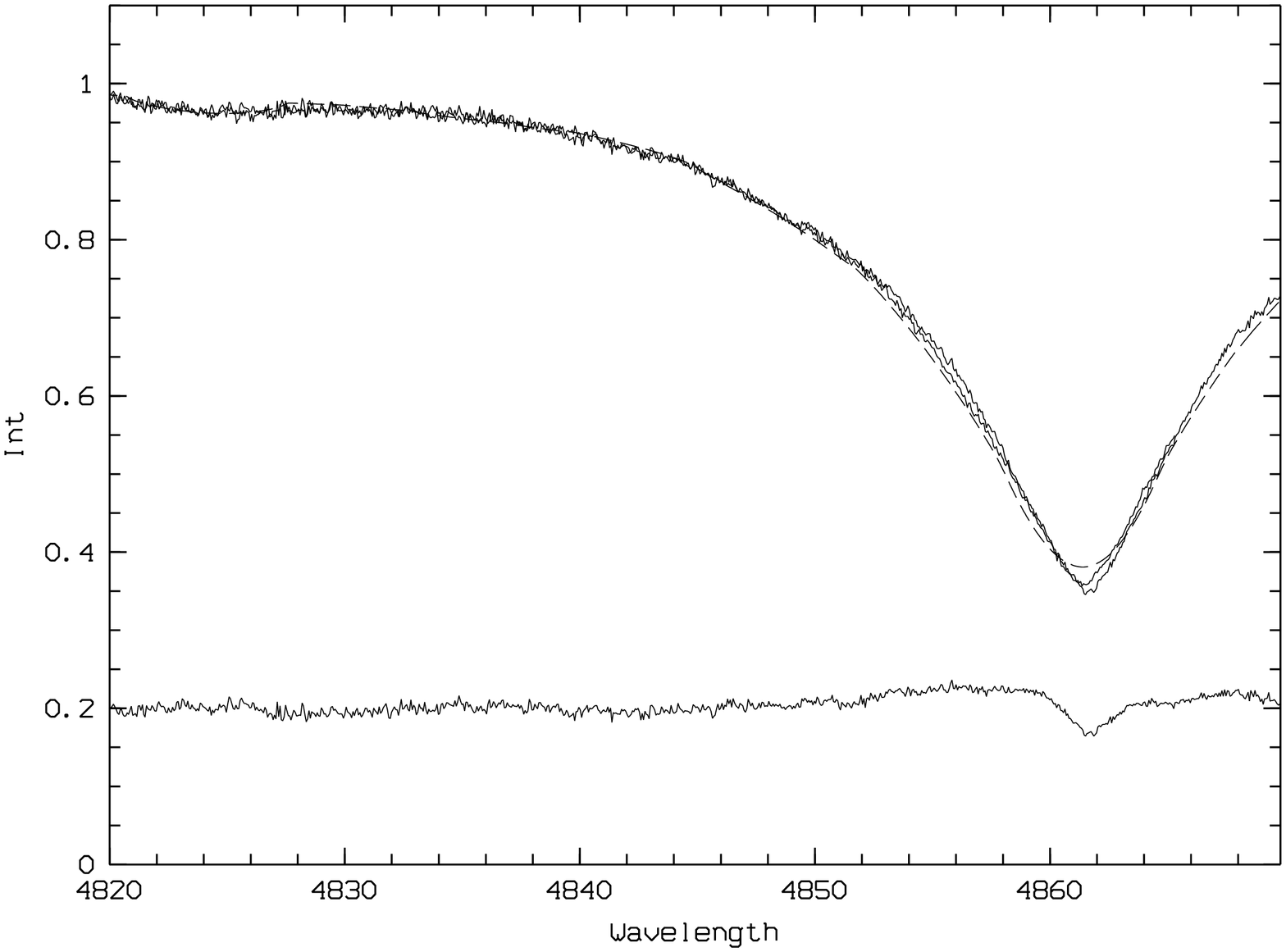,width=8cm,angle=270}}
  \parbox[l]{8cm}{ 
  \caption{H$\alpha$ (top panel) and H$\beta$ (bottom panel) profiles 
  of $\kappa$\,UMa along with the subtracted residual spectrum (observed -- synthetic)
  shifted upward by 0.2 for display purposes.
  Two spectra, obtained in April 2000 and in December 2000, are plotted. 
  The spectra coincide. Solid line~-- observed spectra\ Dashed line~-- synthetic spectrum (see text).
  Weak emission is seen in the H$\alpha$ profile residual, and it is absent in the H$\beta$ profile residual.
  The sharp core of H$\beta$ is likely due to the secondary companion.
  The core is shifted by approximately 20\,km/s toward longer wavelength.
  Similar radial velocity is expected for the secondary from the orbital
  parameters given by Barnaby et al.~(2000).}
  \label{compare}}
\end{figure}

A similar distortion of the line profile which may be due to the presence of
weak emission is seen in H$\alpha$ 
of another A1Vn star from the survey list, $\kappa$\,UMa 
(see Fig.~\ref{compare}). The H$\alpha$ line profile slope changes abruptly
on the flanks of the absorption core, at about $\pm5$\,\AA\,
from the line center.
Ghosh, Apparao and Pukalenthi~(1999) did not detect emission in this star 
during their search for new Be stars among An/Bn stars; however, their survey was undertaken using 
low spectral resolution and they may have therefore missed Ae/Be stars with weak emission.
In our spectra of $\kappa$\,UMa
the H$\alpha$ line core is also much narrower than is expected from
its BSC rotational velocity of 219~km/s.
The star was observed in April, 2000, and again eight months later.
From comparison of the two spectra it appears that the 
emission is rather stable.

Although we suspect that emission is responsible for the peculiar line profiles of $\nu$~Cyg and $\kappa$~UMa, the distortion may also be due to binarity.
In principle, the superposed absorption profiles from two companions in a 
binary system may be mistakenly considered as an emission.
When the observed separation between absorption cores is small 
the profiles are not seen separately, and the combined profile 
may not correspond to any reasonable single-star model.
Binarity often occurs among Ae stars (see Jaschek and Andrillat~1998).
We investigated the literature and found that both two new Ae stars are
indeed binaries. One may suggest that the H$\alpha$ profile 
in Fig.~\ref{compare} simply comprise two profiles that have different width.
If this is the case, we expect other hydrogen lines to have similar shapes.
On the other hand, if emission is responsible for the distortion, its
strength should decrease rapidly to higher order hydrogen lines.
Weak emission appearing at H$\alpha$ should therefore not be found at H$\beta$.
We checked other lines in the spectra of
$\kappa$\,UMa and did not find any signs of the line profile distortion observed in H$\alpha$.
For example, Fig.~\ref{compare} shows the profile of H$\beta$ in the spectrum of $\kappa$~UMa which is free from any profile distortion similar to that 
observed in H$\alpha$. In fact, the H$\beta$ profile is neither fully 
consistent with the synthetic profile superposed in Fig.~\ref{compare}, nor 
is it fully symmetric. 
However, the weakness of these effects relative to 
those found in the H$\alpha$ profile strongly supports our conclusion that 
the H$\alpha$ profile peculiarities are not due to binarity, 
but rather result from emission.

The presence of two stars allows us to derive the dynamical parallax, 
and therefore to estimate the distance to the system. 
$\kappa$\,UMa has an orbital period of $35.61\pm0.06$\,yr and a semimajor 
axis of $185\pm2$\,mas (Barnaby et al.~2000).
Assuming masses of the components to be the same and equal to 3\,$M_{sun}$,
the calculated value of the dynamical parallax is $\approx 9$\,mas.
The trigonometric parallax obtained with HIPPARCOS, $7.7\pm0.83$\,mas 
(Perryman et al.~1997), is  consistent with this value. 
Both results lead to a distance to the system of approximately 110--120\,pc.
According to HIPPARCOS photometry, the visual magnitudes 
of the primary and the secondary are, respectively, $4.16\pm0.11$
and $4.54\pm0.16$. 

$\nu$\,Cyg also has a close companion. According to the HIPPARCOS archive, 
the companion is 2.3\,mag fainter than the Ae primary whose visual magnitude 
is $4.07\pm0.01$.
The orbital parameters of this system are not known. Therefore, the distance to the system, $109\pm6$\,pc, was calculated using only the trigonometric parallax.

Using the calculated distances and visual magnitudes, the 
absolute magnitudes can also be determined. These are
$-1.4$ and $-1.0$ for the primary and the secondary of $\kappa$\,UMa,
respectively, and $-1.3$ and $1.2$ for the primary and the secondary of 
$\nu$\,Cyg.
The absolute magnitudes of the $\kappa$~UMa components and the $\nu$~Cyg primary appear to be almost
one magnitude higher than those of normal A1 dwarfs and correspond to 
those of giants.
Jaschek and Andrillat~(1998) point out that the absolute magnitudes of
Ae and A-type shell stars exceeds the mean value for normal dwarfs.

In order to further confirm the weak emission in H$\alpha$ of $\kappa$\,UMa
we have computed synthetic H$\alpha$ and H$\beta$ line profiles.
The atmospheric parameters used to compute these profiles were derived from several sources.
Malagnini et al.~(1982) obtained the value of $9600$\,K for the combined effective 
temperature of the components of $\kappa$\,UMa by comparison of UV spectra 
with Kurucz models. These authors claim that the value they derived should be  
accurate within 200\,K. 
This is consistent with the temperature we obtain from $uvby\beta$ 
photometry taken from Gray and Garrison~(1987), $T_{\rm eff}=9640\pm200$\,K,
using the calibration by Moon and Dworetsky~(1985). 
From the $uvby\beta$ photometry, we also computed $\log g = 3.3\pm0.1$,
indicative of a giant, and supporting the derived absolute magnitudes.
Although these results are suggestive, we stress that they correspond to the combined spectrum of the $\kappa$~UMa primary and secondary. These conclusions should be robust if the spectral types of the components are similar. 

$\kappa$\,UMa and $\nu$\,Cyg have similar distances and 
visual magnitudes of their primaries. Therefore, both primaries must have
similar absolute magnitudes. Note that the secondary in $\kappa$\,UMa 
must also have the absolute magnitude that is comparable to that of the 
primary (their visual magnitudes are similar).
In the case of $\nu$\,Cyg, the photometry is dominated by the colors of 
the primary - the contribution of the secondary is negligibly small.
The combined colors of $\kappa$\,UMa and $\nu$\,Cyg do not
differ significantly (Gray and Garrison~1987).
This is possible if the primary and the secondary in $\kappa$\,UMa have
similar colors.
Together with similar absolute magnitudes it leads to
similar temperatures and gravities.

Comparison of observed and calculated hydrogen line profiles 
also supports the above conclusion.
Except for the core, which is distorted by emission,
the H$\alpha$ profiles of $\nu$\,Cyg and $\kappa$\,UMa are fully similar.
The difference between the profiles is less than the estimated errors due to improper spectral 
normalization. At H$\alpha$ these errors are found to be less than 1\%.
The same is true for the H$\beta$ profiles, but the agreement is somewhat poorer.
Taking into account that the contribution of the secondary in $\nu$\,Cyg 
is small, one may suggest that both companions in $\kappa$\,UMa
have similar hydrogen line profiles.

Taking into account the apparent similarity of the physical parameters of the components of $\kappa$\,UMa,
we adopted the value 9600\,K for the effective temperature,
3.3 for the logarithmic surface gravity, a microturbulence value equal to 4\,km/s and a solar abundance.

A synthetic spectrum of $\kappa$\,UMa has been calculated using the code 
SPECTRUM developed by Gray (Gray and Corbally~1994).
SPECTRUM uses a Kurucz atmosphere model as input.
The synthetic spectrum has been convolved with the rotational profile 
of desired width and limb darkening factor 
(the latter is assumed to be equal to 0.4). 
We computed synthetic profiles for three lines, H$\alpha$, H$\beta$ 
and MgII~$\lambda$\,4481.
MgII~$\lambda$\,4481 is insensitive to Teff and $\log$\,g in this 
subspectral type, but is sensitive to the projected rotational velocity. 
It is also the only metallic line in our spectra with a measurable profile. 
We use it to determine v$\sin$\,i. 

From the comparison of observed and synthetic profiles of 
the MgII~$\lambda$\,4481 line, we find that v$\sin$\,i is equal to 
$210\pm10$\,km/s. This value is consistent with the value indicated by the BSC, 219\,km/s. $\kappa$\,UMa is clearly a rapid rotator.

Fig.~\ref{compare} presents superimposed observed and synthetic 
H$\alpha$ and H$\beta$ profiles for $\kappa$~UMa. Residual spectra (observed minus synthetic) are also shown.
\begin{figure}
  \vbox{
  \psfig{figure=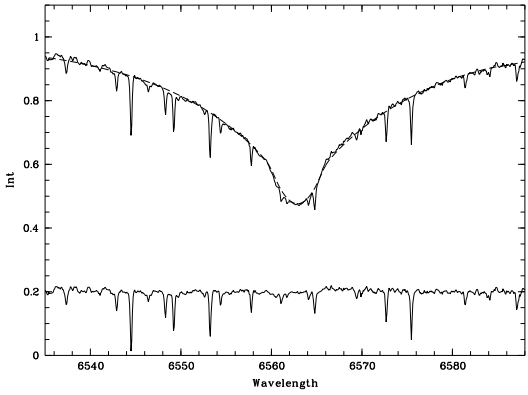,width=8cm,angle=270}}
  \parbox[l]{8cm}{ 
  \caption{Same as Fig.~\ref{compare}, but for the normal A star $\alpha$\,Lac.
  The sharp lines are telluric absorption lines.}
  \label{nonAe}}
\end{figure}
A weak double-peaked emission feature is seen in the residual spectrum well inside the core of 
H$\alpha$. After the subtraction of the synthetic profile, the equivalent 
width of the emission component is 0.3\,\AA. 
In fact, weak emission of comparable strength was detected also in 
the spectra of $\nu$\,Cyg in 1999. The emission strength of $\nu$\,Cyg
decreased by a factor of 2 between 1996 and 1999.

To illustrate that the features seen in the residual spectra are not likely a result of poor determination of the atmospheric parameters (and therefore poor correspondence between the observed and synthetic profiles in the absence of emission),
in Fig.~\ref{nonAe} we present observed, synthetic and subtracted residual
spectra of the normal A star $\alpha$\,Lac (A1V). 
The synthetic spectrum for $\alpha$\,Lac was calculated with the following 
parameters: $T_{\rm eff}=9500$, $\log g=4.0$, and $v\sin i = 130$\,km/s.
The temperature and gravity were estimated from the spectral type.
$v\sin i$ was taken from the BSC.
It is seen from Fig.~\ref{nonAe} that
the mean intensity in the subtracted spectrum of $\alpha$\,Lac
is around zero. The calculated model appears to fit the observed spectrum of a normal A dwarf very well if the spectrum contains no emission.

All measured parameters of the H$\alpha$ emission features 
observed in the spectra of $\kappa$\,UMa and $\nu$\,Cyg 
are presented in Table\ref{EW}.
This table gives the equivalent width of
H$\alpha$ (EW), the equivalent width of the emission ($EW_{em}$)
and the peak-to-peak separation of the emission ($\Delta V_{peak}$).
Note that telluric absorption features were
extracted and did not affect the calculations of the equivalent width.
The equivalent widths of H$\alpha$ reported in Table~\ref{EW} are close to those obtained by Jaschek and
Andrillat~(1998) for Ae and A type shell stars.
\begin{table}
  \caption[]{H$\alpha$ measurements. $EW$ is the equivalent width of H$\alpha$
  (including emission). $EW_{em}$~-- the equivalent width of the emission
  measured in the residual spectrum (see text and Fig.~\ref{compare}).
  $\Delta V_{peak}$ is the peak-to-peak separation of the emission.
  The central position of peaks was measured by fitting a Gaussian.}
  \label{EW}
  \begin{tabular}{rrrrrr}
  \hline
         Name & Date     & JD +    & $EW$, & $EW_{em}$,& $\Delta V_{peak}$,\\
              &          & 2450000 & $\AA$ & $\AA$     & km/s              \\
  \hline                                                  
$\kappa$\,UMa & Apr,2000 & 1658.34 & 10.62 & 0.25      & 294               \\
              & Dec,2000 & 1892.54 & 10.11 & 0.38      & 282               \\
$\nu$\,Cyg    & Aug,1996 &  326.84 &  9.79 & 0.97      & 245               \\
              & Jul,1999 & 1362.50 & 10.68 & 0.51      & 253               \\
  \hline
  \end{tabular}
  \end{table}
$\Delta V_{peak}$ contains information about the size of the emission disk. 
The radius of the disk $R_{disk}$ depends on the rotation law and can be 
calculated from the following equation (Jaschek and Jaschek~1992):
$$\frac{R_{disk}}{R_*} = \left(\frac{2 v\sin i}
{\Delta V_{peak}}\right)^{\gamma}$$
where $R_*$ is the photospheric radius, $\gamma = 1$ for rotation with 
conservation of angular momentum and $\gamma = 2$ for Keplerian rotation. 
Using the data from Table\ref{stars} and Table\ref{EW}
we calculated the radius of the disk.
In case of $\kappa$\,UMa it is found to be between $1.5\pm0.1 R_*$
and $2.1\pm0.4 R_*$ ($\gamma$ is between 1 and 2).
The radius of the disk of $\nu$\,Cyg is between $1.9\pm0.2 R_*$ and
$3.7\pm0.6 R_*$. 
Radii of disks in Ae and Be stars have similar sizes, $2-4 R_*$ 
(Ghosh, Apparao and Pukalenthi~1999; Jaschek and Jaschek~1992).
Thus, the sizes of the disks of these two new Ae stars have 
values typical of Ae/Be stars.

\section{IR properties}

Both $\nu$\,Cyg and $\kappa$\,UMa were detected with IRAS
(see Table\ref{IR}). In the table we present the IRAS number and the stellar 
magnitude in $V$ and $12\mu m$. A color correction was
applied when we calculated broadband magnitudes from the IRAS fluxes.
We followed the color correction scheme presented in the IRAS Explanatory 
Supplement. A color correction factor for a 10000\,K blackbody was assumed.
\begin{table}
  \caption[]{IR data: the visual magnitude, the stellar magnitude at $12\mu m$
  calculated from the IRAS flux, and the $(V-12\mu m)$ color excess.}
  \label{IR}
  \begin{tabular}{rcccr}
  \hline
    Name      &     IRAS   &  V   & [12] & (V-[12])~--  \\
              &            &      &      & (V-[12])$_0$ \\
  \hline                             
$\kappa$\,UMa & 09002+4721 & 3.60 & 3.46 & -0.03        \\
$\nu$\,Cyg    & 20553+4058 & 3.94 & 3.83 &  0.00        \\
  \hline
  \end{tabular}
  \end{table}

A large fraction of A type stars observed by IRAS show infrared excess 
(Jaschek et al.~1991). The IR excess is thought to arise from a circumstellar 
disk. In classical Ae/Be stars the excess is due to free-free emission from
plasma in a gaseous disk. Herbig Ae/Be stars, on the other hand, exhibit IR excess which is
due to thermal emission from hot and/or cool circumstellar dust (Waters and Waelkens~1998). 
These two groups
are well separated on ($V-12\mu m$) color excess versus temperature diagram
(Hillenbrand et al.~1992). In particular, Herbig Ae/Be stars show an IR excess which is
several magnitudes larger than is found for classical Ae/Be stars of a given temperature.
Around 10000\,K classical Ae stars exhibit an excess close to zero, while
it is from 3 to 10\,mag for Herbig Ae/Be stars.

To check whether the new emission stars are classical Ae stars or belong
to Herbig Ae/Be stars, we calculated their ($V-12\mu m$) color excess
(Table\ref{IR}).
Instead of $(V-12\mu m)_0$ we used $(V-N)_0$ tabulated by Johnson~(1966). 
The interstellar extinction for these stars is small at $V$ and negligible at $12\mu m$.

On the ($V-12\mu m$) color excess vs. temperature diagram both 
$\nu$\,Cyg and $\kappa$\,UMa have been found to occupy the same region
with classical Ae stars (with color excesses near 0; see Table~\ref{IR}. It is clear that they are not young 
pre-main sequence Herbig Ae/Be stars.
   
\section{Discussion}

In this study we demonstrated that weak emission ($EW <0.5 \AA$) is observed 
in the H$\alpha$ lines of two new Ae stars, $\nu$~Cyg and $\kappa$~UMa. The presence of weak emission introduces only a small 
distortion in the H$\alpha$ line profile when it is seen in high resolution 
spectra ($R\simeq 40000$) and may be missed in spectra with lower resolution and/or signal-to-noise ratio. 
The distortion of the line profile can be revealed through comparison with a 
synthetic spectrum, provided that the atmospheric parameters of the star are accurately known.

The difficulty of detecting weak emission may result in underestimation
of the fraction of emission-line stars.
The strength of emission decreases with decreasing of temperature.
This suggests that the fraction of stars with weaker emission may be higher for 
late B stars than for early B stars. The frequency of known Be stars also monotonically 
decreases from early to late B types (Zorec and Briot~1997). 
It may well be that effect is at least partly due to the systematic
non-detection of weak emission Be stars.

The group of Ae stars seems to be continuation of Be stars toward
lower temperatures. 
If one continues the observed frequency distribution of Be stars toward early A type stars, one would 
expect approximately 3\% of A1-A2 type dwarfs to be classified as Ae.
Zorec and Briot~(1997) obtained the number of Ae stars by counts in
different catalogues. From their data it appears that only 0.2\% 
of A1-A2 stars have so far been identified as Ae stars. 
This rate is 15 times smaller than is predicted 
from an extrapolation of the distribution.
It therefore seems that there is sharp cutoff in the distribution
near the latest B subspectral types. It is interesting that a sudden change 
in the emission strength also occurs around A0 subspectral type
(Andrillat, Jaschek and Jaschek~1986). 
The emission becomes weaker in early A type stars, and it disappears 
completely in middle A type stars. However, weak emission is harder to detect, and so Ae stars, which are as a rule characterised by weak emission, may be missed more frequently than Be stars. This may lead to an underestimation 
of the frequency of Ae stars. Thus, the real frequency of Ae stars may be substantially higher than that found from simple counts in available catalogues.

\section{Conclusion}

Two new Ae stars, $\nu$\,Cyg and $\kappa$\,UMa, have been discovered 
with the context of 
the Magnetic Survey of Bright Main Sequence stars. They both showed
the presence of emission in the H$\alpha$ line. No significant emission is seen in the H$\beta$ line.
$\nu$\,Cyg exhibits double-peaked emission which is variable on a time scale 
of 3 years.
The H$\alpha$ line profile of $\kappa$\,UMa hosts weak emission.
Its equivalent width is only 0.3\,\AA. Although such weak emission only 
slightly distorts the line profile, we were able to detect this small effect
through comparison of observed and synthetic spectra.
The emission is thought to arise from circumstellar disks, and using the derived parameters of the emission features we estimated the size of the disks.

We derived the absolute magnitudes of the new Ae stars, corrected for binarity.
The Ae stars appear to be almost one magnitude more luminous in $V$ than normal dwarfs
of the same temperature. This supports the conclusion of Jaschek and
Andrillat~(1998) regarding the overbrightness of Ae and A type shell stars.

Both $\nu$\,Cyg and $\kappa$\,UMa are IR sources that were detected
with IRAS. Both exhibit IR fluxes typical of classical Ae stars.
The newly-discovered emission stars are clearly not Herbig Ae/Be stars.

Lack of detection of weak emission similar to that which we have detected 
in $\kappa$\,UMa and in $\nu$\,Cyg in 1999
should be taken into account when calculating the fraction of Ae and Be
stars. This is especially important for Ae stars since they exhibit
weaker emissions than their hotter counterparts.
Counts in presently-existing catalogues may underestimate
the true frequency of Ae stars.

\begin{acknowledgements}
We thank G.G.\,Valyavin and T.A.\,Burlakova for their help and assistance
during observations. We are grateful to J.D.\,Landstreet for his helpful
comments and proofreading of the manuscript.
The VALD database operated in Vienna, Austria and the Simbad database 
operated at the CDS, Strasbourg, France were used during the course of this 
investigation.
This work has been partly supported by the Natural Sciences and Engineering 
Research Council of Canada (NSERC) and partly by RFBR grant N\,01-02-16808.
\end{acknowledgements}

\end{document}